\documentclass[useAMS,usenatbib,usegraphicx]{mn2e}

\newcommand{\grad}{^{\mbox{\small o}}}

\newcommand{\kpc}{\mbox{\,kpc}}

\newcommand{\pc}{\mbox{\,pc}}
\newcommand{\Myr}{\mbox{\,Myr}}

\newcommand{\kms}{\mbox{\,km s}^{-1}}

\newcommand{\pcc}{\mbox{\,cm}^{-3}}
\newcommand{\dif}{\mathrm{d}}
\newcommand{\Msun}{\mbox{\,M}_{\odot}}

\title[]
{Analysis of the spiral structure in a simulated galaxy}

\author[Mata-Ch\'avez, G\'omez \& Puerari]
{Mata-Ch\'avez, M. Dolores,$^1$ \thanks{E-mail: m.mata@crya.unam.mx},
G\'omez, Gilberto C.$^1$, and
Puerari, Iv\^anio$^2$
\\
$^1$Centro de Radioastronom\'ia y Astrof\'isica,
Universidad Nacional Aut\'onoma de M\'exico, Apdo. Postal 3-72,
Morelia Mich. 58089, M\'exico\\
$^2$ Instituto Nacional de Astrof\'isica, \'Optica y Electr\'onica, Calle
Luis Enrique Erro 1, 72840 Santa Mar\'ia Tonantzintla, Puebla,
M\'exico}

\begin{document}
\maketitle

\begin{abstract}

We analyze the spiral structure that results in a numerical
simulation of a galactic disk with stellar and gaseous components
evolving in a  potential  that includes an axisymmetric halo and
bulge. We perform a second simulation without the gas component to
observe how it affects the spiral structure in the disk. To quantify
this, we use a Fourier analysis and obtain values for the pitch
angle and the velocity of the self-excited spiral pattern of the
disk. The results show a tighter spiral in the simulation with
gaseous component.  The spiral structure
is consistent with a superposition of waves, each with a constant
pattern velocity in given radial ranges.

\end{abstract}

\section{INTRODUCTION}

The spiral structure of disk galaxies has been studied for many
years now, yet the origin of this structure remains uncertain.
Several different theories have been proposed to explain how this
structure was formed. The density wave theory \citep{Lin-shu1964,
bertinlin1996} proposes quasi-stationary density waves propagating
through a rotating disk at constant pattern angular velocity. As an
alternative, the Swing Amplification Theory \citep{Goldreich1965,
Julian1966} proposes that the arms arise from smaller perturbs (or
perturbations) which add and amplify. This could produce
over-densities rotating with the disk. In this model waves are not
quasi-stationary and, therefore when the perturbs ceases the spiral
disappears. \cite{D'onghia2013} showed that over-densities could
produce non-linear effects in swing amplifications that modify the
formation and longevity of the spiral pattern, even after the
perturbations have been removed. 

In order to study these scenarios, researchers have   used N-body
simulations.  These have been able to reproduce a spiral like
structure generated in many different ways, such as interaction with
other galaxies or gravitational instabilities in the disk.  Most of
these simulations involve a stellar disk only
\citep{Grand2012,Grand2012a,Roca2013,Quillen2011}, yet the spiral
structure is conspicuous in the gaseous component also
\citep{Acreman2010}.  \citet[and references therein]{Valle2005}
showed that the Galactic stellar spiral structure differs from the
gaseous one \citep[see also][]{Gomez13}.

Recently, \cite{wada2011} and \cite{baba2013} showed  simulations
with stellar and gaseous disks. In those studies, a spiral structure
that seems to co-rotate with the galactic disk is formed.

Several methods have been developed to quantitatively describe the
spiral structure in a galactic disk, either in observed images or in
a numerical simulations. Using Fourier transformations of images of
spiral galaxies, it is possible to obtain estimations of the pitch
angle, relative strengths of modes and other parameters of the
spiral structure \citep{Davis2012}.

But the issue of the gas role in the formation of the spiral pattern
remains unsatisfactorily open.
It is usually considered (\citep{ bertinlin1996,sellwood2014, dobbs2014}, for example) that the principal role of the
 gas is to dynamically cool the stellar disk. Nevertheless, the large scale 
interaction of these components have not been properly explored since it is 
assumed that the small mass of the gaseous disk will have a negligible impact on
 the dynamics of the stellar one, and so the gaseous component is frequently 
considered as a perturbation, brushing aside the possible dynamical feedback on the large scale dynamics.
In this paper, we compare the spiral structure of galactic disks with and without a gaseous component using 3D numerical simulations. We use Fourier
transforms to measure the parameters of the spiral pattern of the
disks.  In \S\ref{sec:method} we describe the simulations performed.
In \S\ref{sec:spiral} we present the analysis of the spiral
structure in the simulations performed. Finally, we present a
summary in \S\ref{sec:sum}.

\section{METHOD}
\label{sec:method}

Our simulation contains a galactic disk with $9.8\times 10^{8}\Msun$
of gas and $3.49\times 10^{10}\Msun$ of stars.  Both components are
distributed initially with a constant midplane density of $0.62$ and
$0.017\Msun/\pc^{3}$ for stars and gas, respectively, out to a
radius of $3 \kpc$.  Outside this radius, the midplane density
follows an exponential profile $\rho = \rho_0 \exp[-(R-R_0)/R_h]$,
with $R_0=8\kpc$, $\rho_0 = 0.15\Msun/\pc^{3}$ and $R_h = 3.5\kpc$
for the stellar disk, and $\rho_0 = 10^{-2}\Msun/pc^{3}$ and $R_h =
8\kpc$ for the gaseous disk.  In the vertical direction, both
components initially follow a Gaussian profile, with scale heights
$0.325$ and $0.135\kpc$ for stars and gas, respectively.

\begin{figure}
  \includegraphics[width=0.5\textwidth]{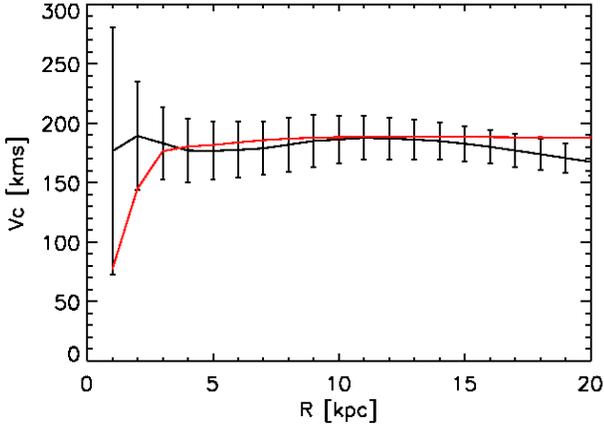}
  \caption{Mean rotational velocity of the simulation. The gray line
    corresponds to $t=0\Myr$ and black line at $t=200\Myr$. The error
    bars show the standard deviation at different radii.}
   \label{fig:veloz}
\end{figure}

The particles in the galactic disk are setup in rotational
equilibrium with a potential similar to that described in
\citet{Allen&Santillan1991}, which consists in a halo, a bulge and a
stellar disk (see fig.  \ref{fig:veloz}).  In addition to the
circular velocity, a velocity dispersion of $20\kms$ is added to the
stars, and of $12\kms$ to the gas.  The Toomre Q parameter for the
disk is $<1$ in the range $2\kpc < R < 7\kpc$.

The simulation is performed with the {\sc GADGET2} code
\citep{Springel2001}, which solves the hydrodynamic equations using
a SPH algorithm coupled to stellar dynamics.  We used
$6\times10^{6}$ stellar particles and $6\times10^{6}$ gas particles,
randomly distributed over the disk following the density profile
described above. The simulations are set up within a $40\kpc$ box.
The version of the {\sc GADGET2} code we use have a sink particle
formation prescription \citep{Jappsen2005} with a critical density
for sink formation of $3\times 10^{3}\pcc$.

Since the mass resolution is similar to the masses of giant
molecular complexes, it is necessary to consider the gas segregation
into phases. To achieve this, the simulation include the cooling
function described in \citet{Koyama2000}.\footnote{ Please note a
typographical error in the expression for the cooling function in
\cite{Koyama2000}. See \citet{Vazquez-Semadeni2007}.}
To avoid a
prohibitively short time step, we apply the fast cooling model
described in \citet{Vazquez-Semadeni2007}, which evolves the gas
temperature as exponentially approaching the equilibrium temperature
at the current density. The segregation of the gas in phases,
the dynamics of the dense clouds formed, and associated star
formation, as modeled by sink particle formation,
will be explored in a future work.

We performed two simulations.  Simulation I consisted on both
stellar and gaseous disk components, while simulation II consisted
only on the stellar component, with the same random density and
velocity distribution as the stellar disk in simulation I.
Simulation I was evolved through $410\Myr$, while we were able to
evolve simulation II through $923\Myr$.
In both simulations, the particles start in an unrelaxed state,
and so the self-stimulated spiral appears sooner than in relaxed
simulations. Nevertheless, since the evolution lasts $3.5
\tau_{rot}$ (where $\tau_{rot}$ is the rotation period at the
stellar disk scale length, $\tau_{rot}=120\Myr$),
the evolution should be enough to erase signatures of the initial conditions.

\begin{figure}
 \begin{center}
  \includegraphics[width=0.47\textwidth]{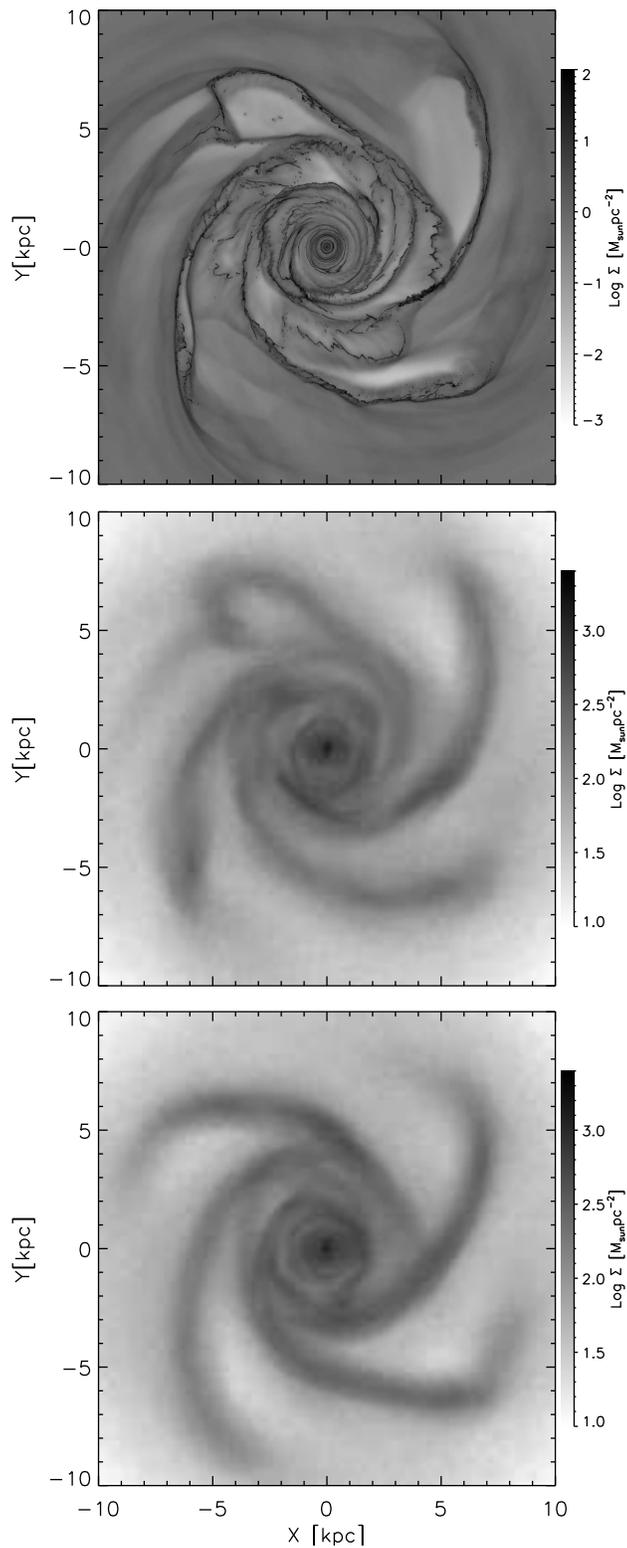}
  \caption{Gaseous ({\it top}) and stellar ({\it middle}) surface density
    distributions for simulation I, and stellar surface density ({\it
    bottom}) for simulation II at $t=200\Myr$.
    For clarity, only the $r < 10\kpc$ is shown.}
   \label{fig:dens}
 \end{center}
\end{figure}

Figure \ref{fig:dens} shows mass surface density maps
of the simulations
at $t=200 \Myr$ for the gaseous disk in simulation I (top) and
stars (middle), and the stellar disk in simulation II (bottom). The
galaxies in the simulations are not perturbed, so the spiral
structure forms due to self-gravity out of the random fluctuations
in the initial particle distribution. In simulation I, the gaseous
and stellar disk have similar large scale structure, namely four
spiral arms with basically the same locus, but the stellar arms are
thicker than the gaseous ones, with the later showing much more
substructure, as expected.  But it is noticeable that the stellar
arms in simulation I show more substructure than those in simulation
II, the most noteworthy being a ``hook'' around $(x,y)=(-4,7)\kpc$,
although the overall strength of the arms differ in both simulations.
The presence of a gaseous component is known to destabilize a disk
\citep{jog1996}, since the gas is dissipative and is allowed to cool.
But, we find that the stellar density in the arms is larger in
simulation II, with the spiral arms remaining coherent longer (see \S
\ref{sec:spiral}).

\section{Spiral structure}
\label{sec:spiral}

\begin{figure}
   \includegraphics[width=0.44\textwidth]{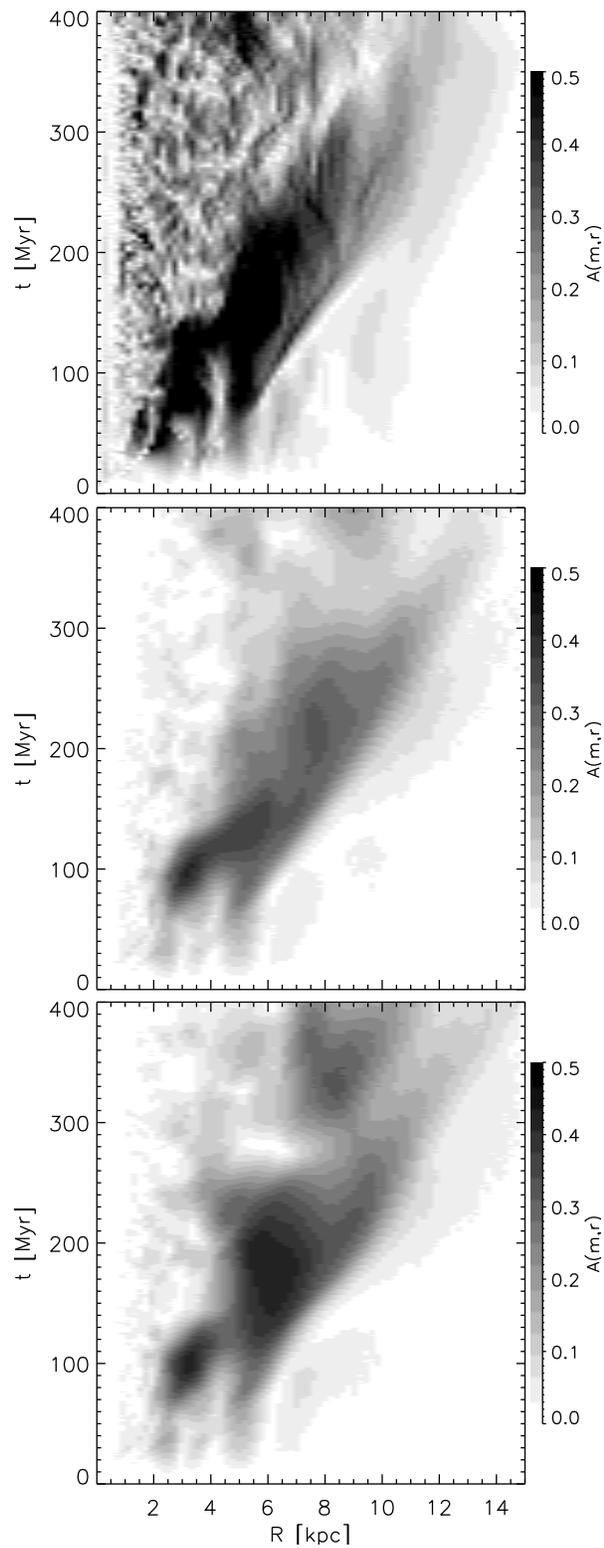}
   \caption{Evolution of the $m=4$ mode the for gas ({\it top}) and stars
     ({\it middle}) in simulation I, and stars ({\it bottom}) in
     simulation II.}
  \label{fig:rvt}
\end{figure}

Following \cite{Grand2012}, we describe the spiral structure in the
simulations using a Fourier analysis of the mass surface density
distributions of stars and gas.  Consider the Fourier transform
(along the azimuthal angle $\phi$, at a given radius $r$) of the
surface density distributions resulting from the simulations,
$A(m,r)$, where $m$ is the Fourier mode in question.
Since both simulations develop four arms (see fig. \ref{fig:dens}), we
show the time evolution of the $A(4,r)$ mode in Figure \ref{fig:rvt}.
It can be
seen that the four-arm structure is formed between $2$ and $5\kpc$
at $t \sim 30\Myr$, extending to larger radii at later times.
But, after a $\sim 300 \Myr$,  while still the largest, the
$m=4$ mode is no longer dominant since other modes grow in the inner
part of the galaxy. After $t \sim 150\Myr$, $A(4,r)$ is
significant only in the $5\kpc < R < 10\kpc$ range, with its
amplitude declining in time.

  It is noticeable that the gas component in simulation I has the
largest amplitudes, meaning that the gas is more tightly associated with a
four-fold symmetric pattern than the stars.
Comparing the stellar density in both simulations I and II,
we may note that simulation II has larger $A(m,r)$ amplitudes,
meaning that in the absence of gas, the spiral
structure in enhanced.
One possible reason for this is that
interactions between stars and giant molecular clouds 
heats the stellar disk, thus causing the spiral structure
intensity to decrease.
Since we failed to find a correlation between the stellar velocity dispersion
and the gaseous disk surface density (for constant radius rings), we do not
think this is the reason for a weaker spiral when gas is included.
Another possible reason is that
the spiral structure in simulation I
is formed almost at the same time in the stellar and gaseous disks, but with a
small phase difference
(\citealt{shu1973}, P\'erez-Villegas, G\'omez \& Pichardo in prep.).
Even if the gaseous arm is small compared to the stellar arm mass, this
out-of-phase perturbation might
cause a decrease on the stellar response.

In order to measure the pitch angle for each of the spiral modes
that describe the density distribution, consider the amplitudes for
the unwound modes,

\begin{equation}
  \hat{A}(m,p) =\frac{1}{D} \int_{-\pi}^{+\pi}\int_{u_{min}}^{u_{max}} \Sigma(u,\phi)\
          \exp\left[-i(m\phi + pu\right)] \dif\phi\,\dif u,
\end{equation}

\noindent
where, $u=\log r$, $m$ is the Fourier mode in question,
$p = -m/\tan(\alpha)$ is a logarithmic wavenumber,
$\alpha$ is the pitch angle of the spiral,
$\Sigma$ is the mass surface
density distribution, and $D$ is a normalization given by

\begin{equation}
D = \int_{-\pi}^{+\pi} \int_{r_{min}}^{r_{max}}\Sigma(u,\phi)
\dif u \,\dif\phi.
\end{equation}

\begin{figure*}
   \includegraphics[width=0.33\textwidth]{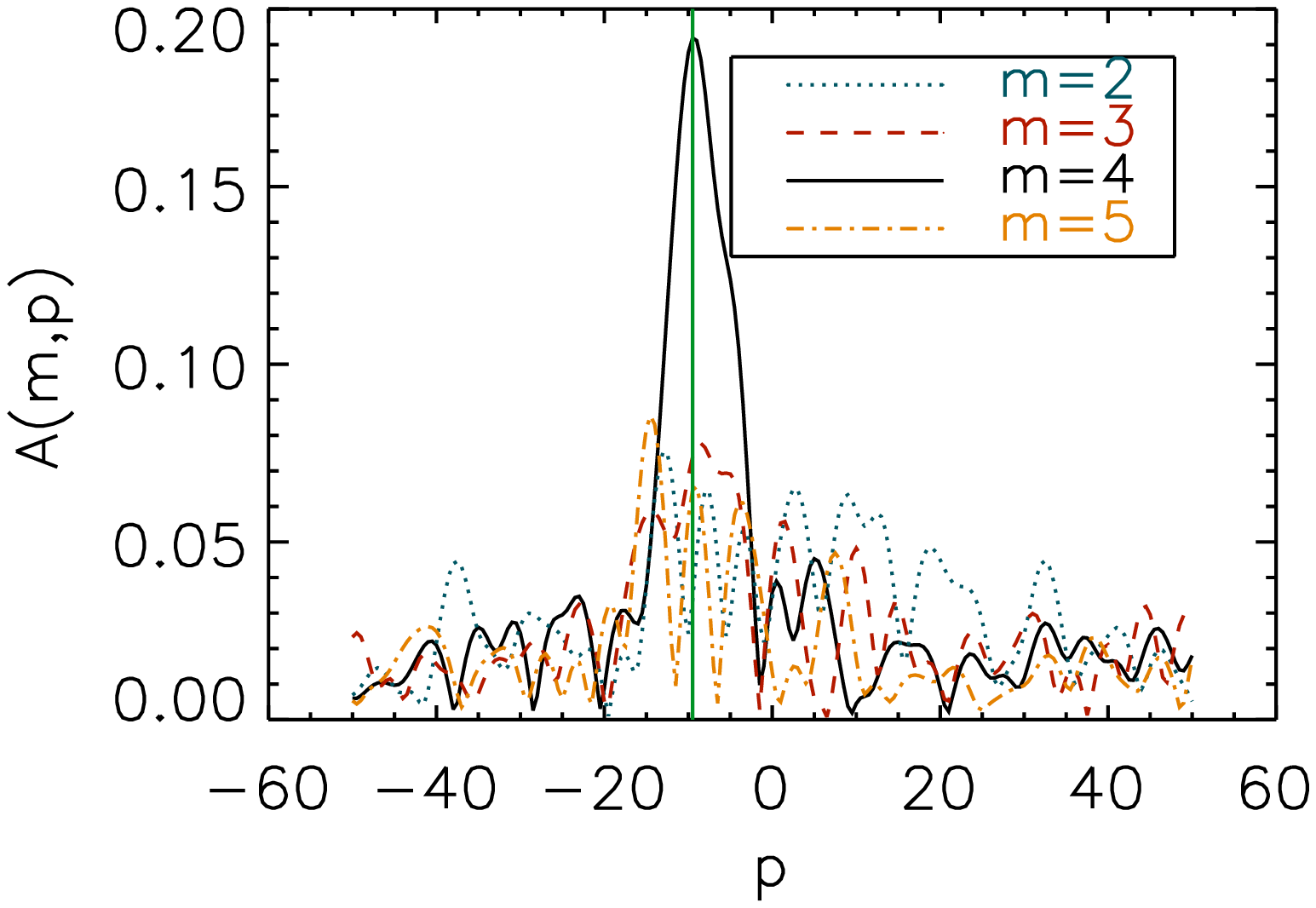}
   \includegraphics[width=0.33\textwidth]{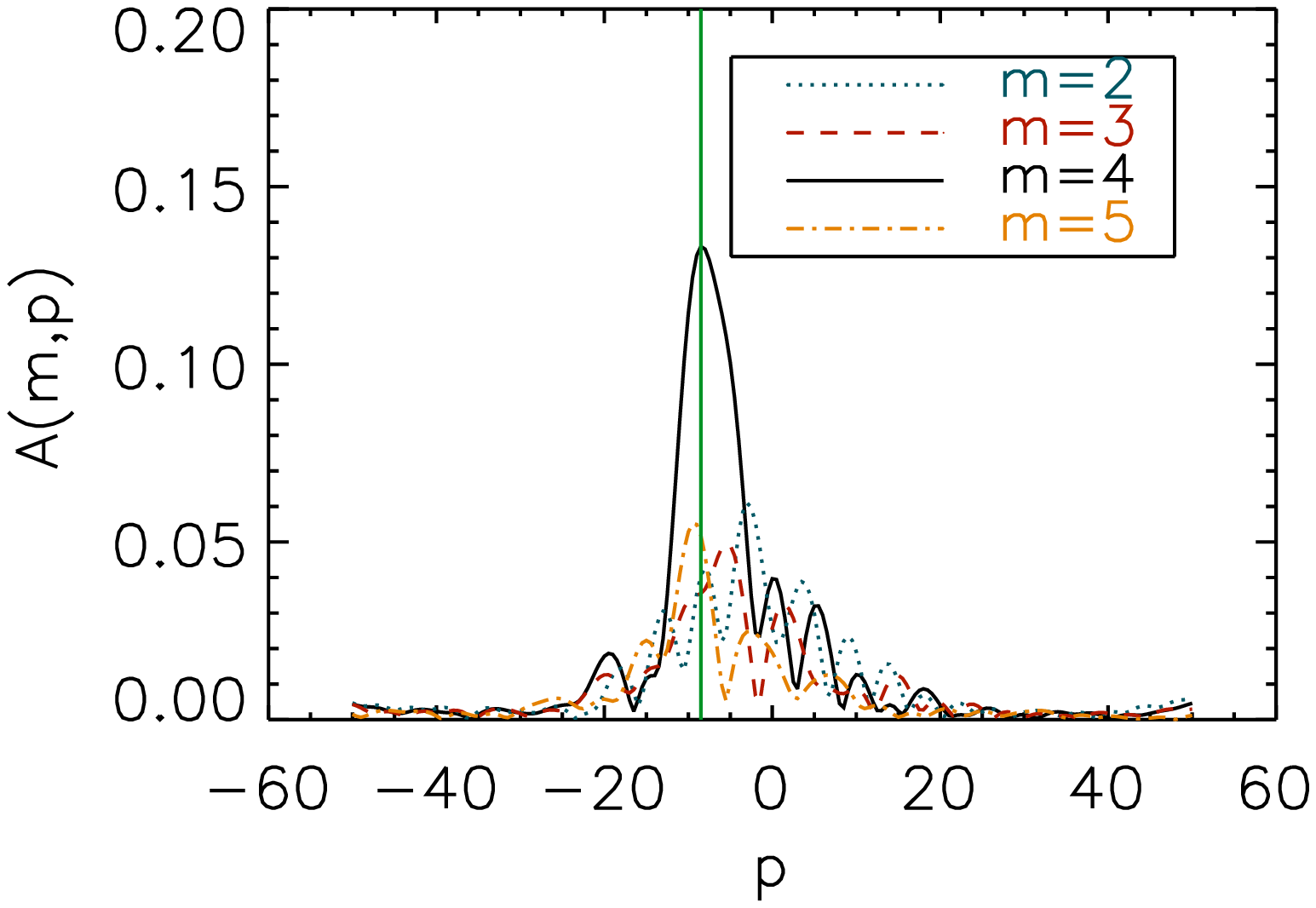}
   \includegraphics[width=0.33\textwidth]{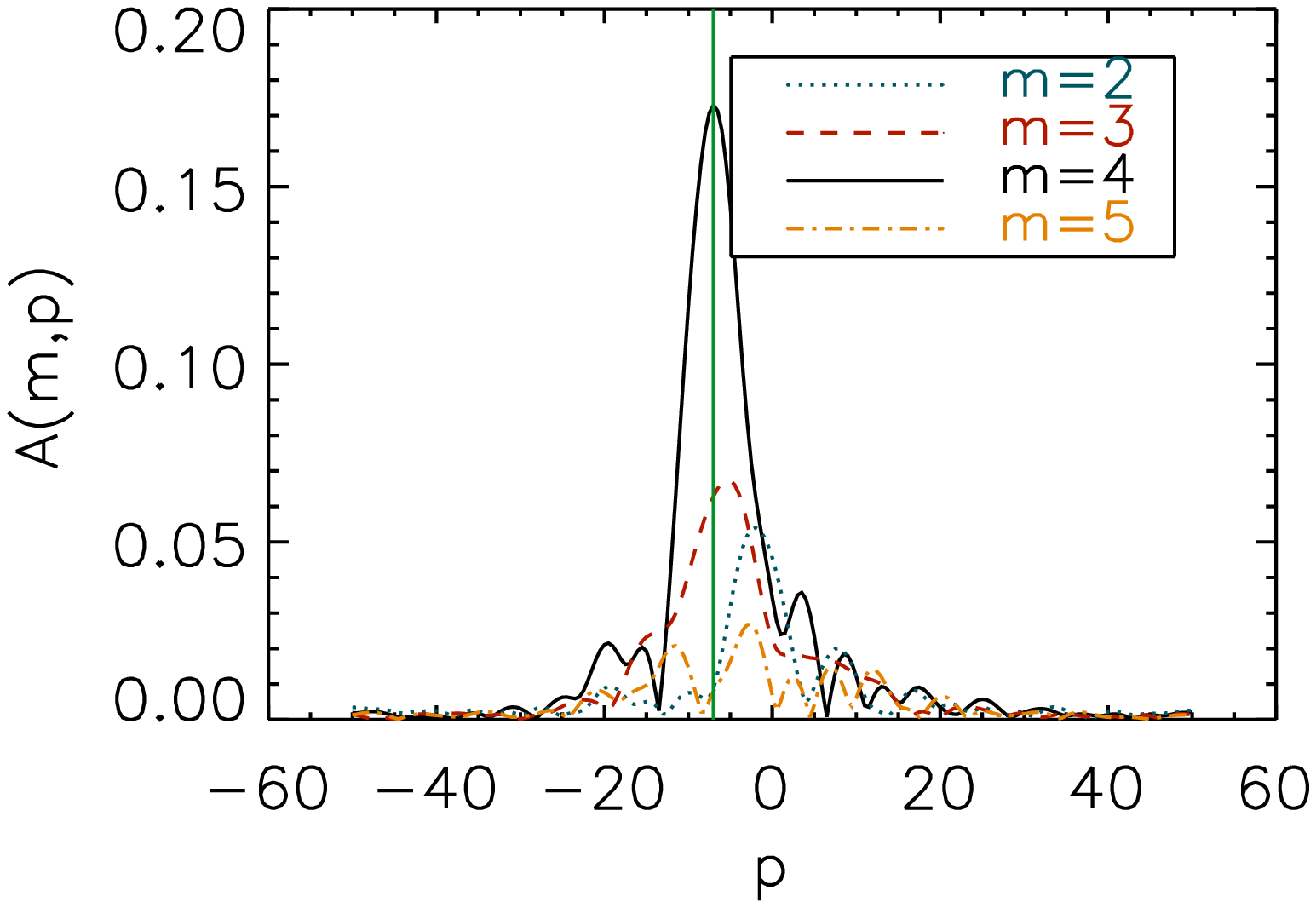}
   \caption{$\hat{A}(m,p)$ for gaseous disk for $t=200\Myr$ ({\it left}),
     for  stellar disk $t=200\Myr$ ({\it center}) and stellar disk
     without gas({\it right}).
     Lines represent different Fourier
     modes.
     The peak of the dominant $m=4$ mode changes from
     $p$-values corresponding to pitch angles of  $\alpha=22.8\grad$
     ({\it left}),  $25.2\grad$ ({\it center}) and $29.7\grad$ ({\it
     right}).  }
   \label{fig:m4}
\end{figure*}

\noindent
Figure \ref{fig:m4} shows the amplitude $\hat{A}(m,p)$  for several $m$-values
for both simulations.  The $m = 4$ mode dominates the distribution
for most of the evolution, and so, hence forth, we focus our
analysis on this mode.

\begin{figure*}
 \begin{center}
   \includegraphics[width=1.01\textwidth]{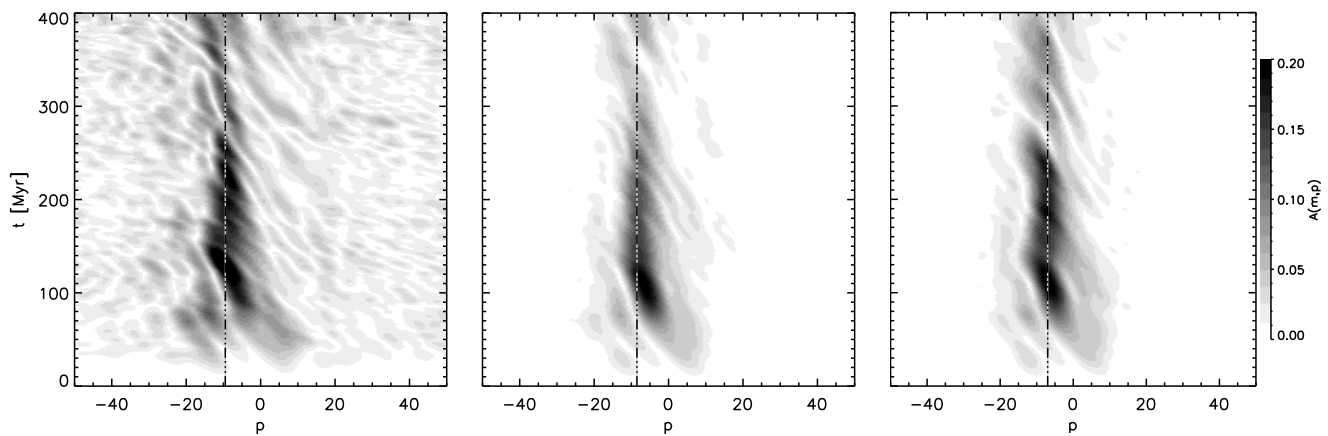}
   \caption{Time evolution of the logarithmic wavenumber, $p$.
     The maximum at
     each time indicate the pitch angle for the gaseous 
     ({\it left}) and stellar disks ({\it center})
     in simulation I, and the
     stellar disk in simulation II ({\it right}).}
\label{fig:pvt}
 \end{center}
\end{figure*}

Figure \ref{fig:pvt} shows the evolution of $\hat{A}(4,p)$ during the
simulation.
In the plot it can be seen that, even if $p$ remains almost constant, the
amplitude $\hat{A}(m,p)$ changes in time.
This might be explained if the spiral is formed by a superposition of
transient waves that reinforce the pattern, as stated by
\citet{sellwood2014}, {\bf or as a result of interference of longer-lived spiral waves, as proposed by \citet{comparetta2012}.}
For simulation I at $200\Myr$, the stellar
spiral is more open ($\alpha=25.2^{\circ}$) than the gaseous one
($\alpha=22.8^{\circ}$), in a similar fashion to simulations of gas
in a fixed potential \citep{Gomez13}.  For simulation II, the pitch
angle at $t=200\Myr$ is $29.7^{\circ}$, i.e, the simulation without
a gaseous disk yields a more open spiral than the simulation with
stars only.  This values are consistent with a Sc galaxy
\citep{Ma2002}.

\begin{figure}
   \includegraphics[width=0.5\textwidth]{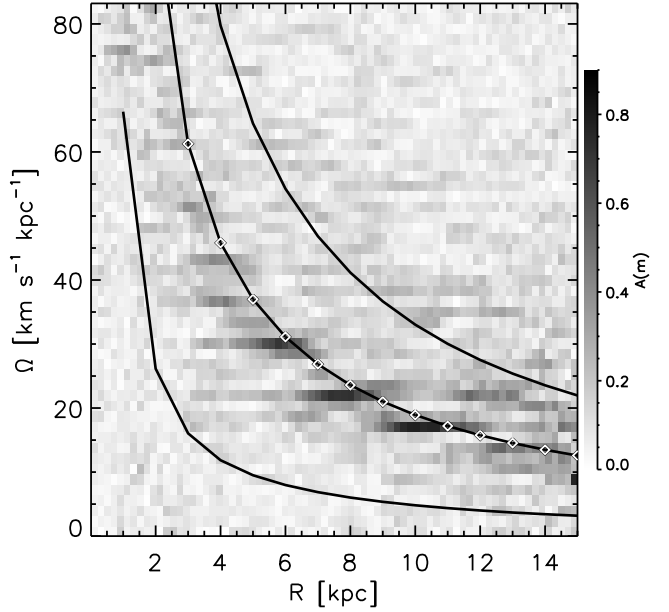}
   \caption{Spectrogram for the $m=4$ mode in the stellar disk of
    simulation II for the length of the simulation, i.e.,
    $923\Myr$.
    The line with diamonds corresponds to
    the angular frequency for the axisymmetric disk, while
    the upper and lower lines represent the $\Omega \pm \kappa/4$
    frequencies.}
 \label{fig:omega}
\end{figure}

In order to determine the velocity of the spiral pattern, consider
the amplitudes of the Fourier modes transformed again in time, as a
function of radius, thus changing from $(m,r,t)$ space to
$(m,r,\Omega_{p})$. Consider the phase $ \Phi =
\arctan(A_{Im}/A_{Re})$ where $A_{Im}$ and $A_{Re}$ are the
imaginary and the real parts of the amplitude. The pattern angular
velocity  is then given by $\Omega_{p} = \hat{\Phi}/N $, where
$\hat{\Phi}$ is the Fourier transform in time of $\Phi$ and $N$ is
the total number of data outputs.

Figure \ref{fig:omega} shows the so calculated spectrogram for the
$m=4$ component, along with the orbital frequency $\Omega$ of the
(initial) axisymmetric disk.
It is worth noting that, since the frequency resolution (the
Nyquist frequency) is
given by the time of the last data output, it is necessary to
allow for the longest possible evolution of the simulation.
In our case, Simulation I stopped due to numerical issues, but the
Simulation II ran up to $923\Myr$.
So, the figure shows the spectrogram corresponding to the
simulation with the stellar disk only, with $1\Myr$ between
data outputs.
As seen from the radial dependence of the amplitude maxima, the
frequency for the spiral pattern is not constant
but it is composed of a superposition of patterns with different
frequencies constant on \emph{restricted} radial ranges, in a manner
consistent to the behavior reported by \cite{sellwood2014}.
This superposition of waves is also suggested in the time
evolution shown in Figure \ref{fig:pvt}.
If Simulation I is used to measure the pattern speed, either with
the stellar or gaseous disks, the lower
frequency resolution smears these constant $\Omega_p$ regions,
giving the impression that the spiral arms rotate solidly,
similarly to those reported recently
\citep{Roca2013, Grand2012a,wada2004,wada2011}.

\section{Summary}
\label{sec:sum}

In this paper we performed SPH simulations of galactic disks using
an SPH-Nbody code to model a disk with a gaseous and stellar
components (simulation I) and a disk with stellar component only (II).
The gaseous disk is modeled with an explicit cooling function, thus
allowing it to segregate into dense and diffuse phases.
We observed  similar 4-arms structure in both simulations but,
when a Fourier analysis is performed on the surface density
distribution,
the spiral structure in simulation II shows a higher $m=4$ mode amplitude,
with less substructure than simulation I, i.e., adding a gaseous
component to the simulation leads to more substructure in both
stellar and gaseous arms, but it also leads to a weaker stellar
spiral.
We speculate that this might be due to a phase shift between the
gaseous and stellar arms, which reduces the coherence of the
response to the non-axisymmetric part of the potential.
This phase shift between stellar and gaseous spiral arms has been
reported before in simulations with fixed spiral potentials
\citep[e.g.][]{shu1973,Gomez13}.

The simulations obtained were analyzed with a Fourier method to
measure the pitch angle and the velocity of the spiral pattern.
The spectrogram for simulation II shows that the pattern is better
described as a superposition
of waves, each with a constant pattern speed in a given radial
range.
Lack of frequency resolution smears the spectrogram and might make
it appear as if the spiral pattern corotates with the disk, as the
set of waves, as a whole, follows the rotation of the disk.

About the growth of spiral pattern, Figure 3 shows the evolution of the spiral structure in our simulations. It can be seen that, even if the spiral structure consists on small fluctuations, it grows globally in a coherent way.
\citet{D'onghia2013} shows that disconnected perturbations serve as seed for the growth of a global spiral pattern through swing amplification, which is favored by the particles' self-gravity.
In a similar way, over-densities in our simulations generate spiral segments that connect and form a large-scale spiral pattern.
The spiral consists on individual segments that rotate with distinct frequencies (as seen in fig. 6), but still a single global pattern emerge.

With respect to the pitch angles of the pattern, we measure
$\alpha=22.8\grad$ for the gaseous disk,
$\alpha=25.2\grad$ for the stellar disk in simulation I and $\alpha
=29.7\grad$ for the stellar disk in simulation II.
A gaseous spiral tighter than the stellar one has been
reported in simulations before.
But, the fact that the stellar spiral develops a larger pitch angle
when the gas is absent appears counter-intuitive considering that the
pitch angle is usually more open for disk galaxies of later Hubble
type, which have a larger gas content.
Further experiments with a range of structural parameters (namely
bulge/disk mass ratio or disk/halo scale length ratio, for example)
are necessary to explore the different ways the stellar and gaseous
disks generate spiral structure in isolated galaxies.

\section*{Acknowledgments}
The authors wish to thank V. Debattista, E. D'Onghia,
A. P\'erez-Villegas, and J. Sellwood, 
for useful discussions on the subject at hand
and an anonymous referee for comments that greatly improved this manuscript.
I.P. thanks the Mexican Foundation CONACyT for finantial support.
The numerical simulations were performed in the cluster at CRyA-UNAM
acquired with CONACyT grants 36571-E and 47366-F to E. V\'azquez-Semadeni.
This work has received financial support from
UNAM-DGAPA PAPIIT grant IN111313 to GCG.

\label{lastpage}
\bibliographystyle{mn2e}
\bibliography{biblio}
\end{document}